\documentclass[reprint,superscriptaddress,footinbib,floatfix]{revtex4-1}
\usepackage{graphicx,color, bm, amsmath, amssymb,hyperref,braket, lipsum, textcomp, mathtools}
\usepackage[dvipsnames]{xcolor}
\usepackage[normalem]{ulem}
\usepackage[english]{babel}
\usepackage{subfigure}

\newcommand{\BE}[0]{\begin{equation}}
\newcommand{\EE}[0]{\end{equation}}
\newcommand{\ketvec}[1]{|#1\rangle\rangle}
\newcommand{\bravec}[1]{\langle\langle#1|}
\newcommand{\braketvec}[1]{\braket{\braket{#1}}}

\newcommand{\up}{\uparrow}
\newcommand{\down}{\downarrow}

\begin{document}
	\title{Photonic cross-noise spectroscopy of Majorana bound states} 
	\author{L. Bittermann}
	\affiliation{Institut f\"ur Mathematische Physik, Technische Universit\"at Braunschweig, D-38106 Braunschweig, Germany}
	\author{F. \surname{Dominguez}}
	\affiliation{Institut f\"ur Mathematische Physik, Technische Universit\"at Braunschweig, D-38106 Braunschweig, Germany}
	\author{P. Recher}
	\affiliation{Institut f\"ur Mathematische Physik, Technische Universit\"at Braunschweig, D-38106 Braunschweig, Germany}
	\affiliation{Laboratory for Emerging Nanometrology Braunschweig, D-38106 Braunschweig, Germany}
	
	\date{\today}

\begin{abstract}
We propose a route to detect Majorana bound states (MBSs) by coupling a topological superconductor to quantum dots (QDs) in a $pnp$ junction. Here, two MBSs are coherently coupled to electrons on two QDs, which recombine with holes to photons. We focus on the spectroscopy of cross-correlated shot noise and the polarization of the emitted photons. Our detection scheme allows us to probe the necessary condition for the emergence of MBSs, specifically, the existence of nonlocal triplet superconducting correlations and also the fundamental property that two MBSs comprise a single complex fermion. We compare our results to the ones obtained from nontopological quasi-MBSs (qMBSs) and establish a correspondence between the number of peaks in the cross-correlation with the number of MBSs in the system. Here, we can identify a tunneling regime that facilitates differentiation between topological MBSs and trivial qMBSs. Additionally, we test the robustness of the detection scheme by the addition of uncorrelated quasiparticles.
\end{abstract}
	
	\maketitle
	
\section{Introduction}

In condensed matter physics, Majorana bound states (MBSs) are unique quasiparticles with neutral charge and an undefined occupation number \cite{Read2000, Ivanov2001, Kitaev2001}. They emerge as zero-energy excitations in topological superconductors (TSCs), which establish themselves as bound states at boundaries or in vortex cores.
Particularly, they can be engineered by proximitizing semiconductor systems with strong spin-orbit interactions and in the presence of a Zeeman field \cite{Fu2008a, Lutchyn2010, Oreg2010}. 
Their interest lies not only at a fundamental level, but also for being building blocks of fault-tolerant quantum computation \cite{Ivanov2001, Alicea2012, Kitaev2003,Nayak2008,Beenakker2020}. 
For more extended reviews, see \cite{Nayak2008, Alicea2012, Beenakker2013a, Aguado2017, Schuray2020, Prada2020, Flensberg2021a}.

There is a plethora of detection schemes that can in principle probe the presence of these exotic quasiparticles. For example, the quantization of the zero bias conductance in a NS junction \cite{Law2009, Prada2012, Li2012} or the fractional Josephson effect \cite{Kitaev2001, Kwon2004}, which can be measured in the Shapiro experiment via the disappearance of the odd Shapiro steps or in the Josephson radiation via the presence of a fractional frequency emission line. Unfortunately, the experimental results differ from ideal predictions, showing a non-quantized value of the zero-bias conductance that ranges from low conductance \cite{Mourik2012, Deng2012, Das2012, Nichele2017} to close to the quantized value \cite{Zhang2019, Yu2021, Wang2022}. 
Also, in the fractional Josephson effect, odd Shapiro steps \cite{Rokhinson2012, Wiedenmann2016, Bocquillon2016,Fischer2022a}
and integer emission line frequencies \cite{Deacon2017a} appear for a finite range in the parameter space.
A plausible interpretation of these results can be explained in terms of the presence of quasi-MBSs (qMBSs) \cite{Kells2012a, Prada2012, Fleckenstein2018, Moore2018, Vuik2019,Marra2022}, appearing naturally in NS junctions, or nonadiabatic transitions \cite{Yeyati2003,San-Jose2012,Dominguez2012a,Pikulin2011a,Houzet2013a,Virtanen2013,Matthews2014a, Sau2017a,Dominguez2017}. Due to this uncertainty, a huge effort has been put forward to distinguish MBSs from trivial excitations, studying theoretically \cite{Nilsson2008,Lue2012,Zocher2013,Rosdahl2018,Pan2021,Pikulin2021,Hess2021,Hess2023} and experimentally \cite{Yu2021, Aghaee2023} their nonlocal nature via nonlocal transport, and measuring the spin-symmetry of the pairing amplitude \cite{Bordoloi2022,Dvir2023,Bordin2023}. 
Other proposals suggest to use local measurements \cite{Ricco2019,Dourado2024} or to study more specific properties, such as their triplet correlations \cite{Haim2015, Fleckenstein2020a,Dominguez2024}.
Furthermore, MBSs were investigated by optical means by coupling MBSs to microwave photons \cite{Ohm2014,Schmidt2013a,Trif2012,Schmidt2013b,Cottet2013,Ohm2015,Dmytruk2015,Dartiailh2017,Contamin2021,Trif2019a,Trif2019b}. Another related idea is the coupling of a Majorana nanowire to a quantum dot (QD) embedded within a microwave cavity investigating the nonlocality of the MBSs \cite{Ricco2022}.

In this manuscript, we study the coherent coupling of MBSs to optically active QDs embedded within a $pn$ junction. Such hybrid systems that combine semiconductor optics and superconductivity were investigated theoretically  \cite{Suemune2006,Recher2010,Hassler2010,Godschalk2011,Baireuther2014,Nigg2015,Schroer2015,Hlobil2015} and experimentally realized \cite{Sasakura2011,Panna2018,Yang2021}, especially the embedment of QDs in $pn$ junctions has been achieved \cite{Minot2007,Versteegh2014}. Similarly to recent charge transport measurements \cite{Bordoloi2022,Dvir2023}, the proposed setup 
allows to read off the spin-dependent superconducting correlations present in the TSC. However, in contrast to those transport setups, here, the spin-information is directly extracted from the polarization of the emitted photons. In our previous work in Ref. \cite{Bittermann2022}, we investigated various signatures of MBSs using a single optically active QD, but could not determine whether these excitations were topologically nontrivial, as local probes are inadequate for this purpose \cite{Prada2020} (an exception being the probe of rotating MBSs \cite{Park2020}). In this work, we have added another QD to have access to nonlocal correlations that are necessary for distinguishing MBSs from trivial excitations.

\begin{figure}[tb]
	\includegraphics[width=\columnwidth]{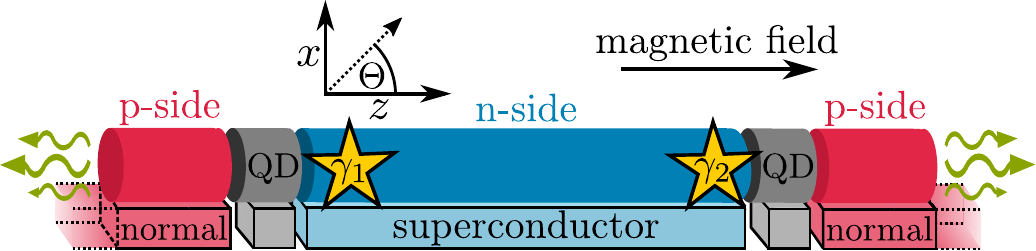}
	\caption{Sketch of a $pnp$ junction to probe MBSs. The $n$ side is a TSC where MBSs emerge at the ends in the topologically nontrivial phase. The $p$ side is coupled to a normal conducting reservoir. In the two $pn$ junctions, QDs are formed that emit photons via electron-hole recombination. As indicated in the figure, a possible realization of this junction is a semiconducting nanowire that can host MBSs when contacted to an $s$-wave superconductor in a magnetic field in $z$ direction. Thus, the Majorana angle $\Theta$ lies in the $x$-$z$ plane. Photons are emitted from the two QDs along the wire direction.}
	\label{fig:setup}
\end{figure}
In particular, we calculate the photonic cross-correlated noise of a $pnp$ junction and highlight the spectroscopic features that are specific for the presence of MBSs. The central part ($n$) consists of a one-dimensional (1D) TSC featuring MBSs at its ends, which are coherently tunnel-coupled to electrons on QDs placed laterally on each side of the TSC. When electrons populate the QDs, they recombine with holes provided from a normal conducting part ($p$), yielding the emission of a polarized photon, see Fig.~\ref{fig:setup}. In this scenario, the presence of nonlocal superconducting correlations on the TSC allows for the correlated emission of one photon on each QD. Thus, we propose to measure the nonlocal or cross-correlated noise of the photon emission processes, which is calculated by studying the full counting statistics of a Markovian master equation, which considers the coupling to the photon bath and hole reservoirs as a perturbation.
We show that the cross-noise is robust to the presence of uncorrelated quasiparticles that can populate the QDs. Moreover, we compare the resulting cross-correlations from MBSs to those of qMBSs, where we observe several features to discern both cases with our detection scheme.

The paper is organized in the following way. In Secs.~\ref{sec:model} and~\ref{sec:master} we introduce the Hamiltonian and master equation of the system, respectively. Then, in Sec.~\ref{sec:correlations} we show the numerical results of the cross-correlations that probe the presence of MBSs and compare these results with the ones provided by qMBSs. In Sec.~\ref{sec:conclusion}, we provide the conclusion. More technical aspects are presented in several Appendices.

\section{Model}\label{sec:model}

We consider a $pnp$ setup shown in Fig.~\ref{fig:setup}. Here, the $n$ side with chemical potential $\mu_n=\mu_e$ is a TSC with a pair of MBSs at the ends. Moreover, on each side of the TSC, the $pn$ junctions contain QDs which have discrete levels for electrons and holes separated by a bias voltage $\mu_n-\mu_p=eV$.
In this scenario, when electrons from the grounded superconductor tunnel coherently into electronic levels of the QDs, they can emit a photon after an electron-hole recombination process occurs.
The process repeats due to the out of equilibrium situation produced by the coupling to a normal conducting reservoir ($p$ side) with chemical potential $\mu_h=-\mu_p$, which provides holes on the QDs. We describe the system by means of the Hamiltonian $H=H_e+H_h+H_{\rm res}+V_h+H_{\rm ph}+V_{\rm rec}$, where $H_e$ describes the coupling between the QD electrons and the MBSs
\begin{align}
\begin{split}
\label{eq:H_e}
H_e&=\sum_{D,\sigma}(\varepsilon_{eD}+\sigma\Delta_Z) d_{D\sigma}^\dagger d_{D\sigma}
+\sum_{D} U_{eD} \hat{n}_{dD\up} \hat{n}_{dD\down}\\
&+\frac{i}{2}\xi \gamma_A \gamma_B +\sum_{D,\sigma}\left(t_{D\sigma} d_{D\sigma} \gamma_D+ \text{H.c.}\right),
\end{split}
\end{align}
where $d_{D\sigma}^{(\dagger)}$ annihilates (creates) an electron with energy $\varepsilon_{eD}+\sigma\Delta_{Z}$ on QD $D=A,B$, spin $\sigma=\up,\down$, and Zeeman energy $\Delta_Z$ having a $z$ component of the total angular momentum of $j_{z,e\sigma}=\pm\hbar/2$. The second term accounts for the intradot Coulomb repulsion, which suppresses double occupation since $U_{eD}\gg \Delta_Z,\varepsilon_{eD},t_{D\sigma},\xi$, where $\hat{n}_{dD\sigma}=d^\dagger_{D\sigma}d_{D\sigma}$ is the occupation number operator \footnote{For the calculations, we use a charging energy of $U_{eD}=100\Delta_Z$.}. Additionally, we assume $U_{eD}\gg\Delta_S$ with the superconducting pairing potential $\Delta_S$, so the usual proximity effect on the QDs is suppressed. The third term describes the coupling between the MBSs $\gamma_{A}$ and $\gamma_B$ with the amplitude $\xi$. The last term accounts for the coupling between the electrons of the QD and MBSs with tunneling amplitudes $t_{D\sigma}$ for MBS $\gamma_D$ and electrons on QD $D$ with spin $\sigma$. Here, the spin subindex of the tunneling amplitudes is determined by the relative spin direction of the electrons in the QD and the electronic components of the Majorana spinor wave function.

We rewrite Eq.~\eqref{eq:H_e} in the basis of the nonlocal fermion $c^\dagger$, which comprises two MBSs $\gamma_A=c^\dagger+c$ and $\gamma_B=i(c^\dagger-c)$, yielding
\begin{align}\label{eq:H_e,fermion}
\begin{split}
&H_e=\sum_{D,\sigma}(\varepsilon_{eD}+\sigma\Delta_Z) d_{D\sigma}^\dagger d_{D\sigma}
+\xi\left(\hat{n}_c-\frac{1}{2}\right)\\
&+\sum_{\sigma}[t_{A\sigma} d_{A\sigma}(c^\dagger +c)
 +it_{B\sigma}d_{B\sigma}(c^\dagger -c) +\text{H.c.}],
 \end{split}
\end{align}
where $\hat{n}_c=c^\dagger c$ is the occupation number operator for the nonlocal fermion. 

We diagonalize Eq.~\eqref{eq:H_e,fermion} in the product basis $\ket{\hat n_{dA\sigma}}\times \ket{\hat n_c} \times \ket{\hat n_{dB\sigma}}$. Here the fermion parity, defined by the sum of occupation numbers in the QDs and wire $(n_{dA\sigma}+n_c+n_{dB\sigma})$ mod~2, is conserved. Thus, the eigenstates $\ket{\psi}_e$ of $H_e$ can be divided into decoupled even- and odd-parity subspaces.

A possible realization of the phenomenological model, introduced in Eqs.~\eqref{eq:H_e} and \eqref{eq:H_e,fermion}, is a semiconducting nanowire with Rashba spin-orbit interaction perpendicular to the wire axis and a magnetic field in $z$-direction along the wire axis \cite{Lutchyn2010, Oreg2010}, where the parameters $\xi$ and $t_{D\sigma}$ can be obtained from.
In this case, the Majorana spin lies in the $x$-$z$ plane, so that we can parametrize it by a single angle $\Theta$, see Fig.~\ref{fig:setup}. Furthermore, the spin polarization of the MBSs on both ends is correlated \cite{Sticlet2012,DasSarma2012,Prada2017}, so that we can describe the tunneling amplitudes by
\begin{align}
\begin{split}
&t_{A\uparrow}=-i t_{B\uparrow}=t\cos(\Theta/2),\\
&t_{A\downarrow}=i t_{B\downarrow}=t\sin(\Theta/2),
\end{split}
\end{align}
with the tunneling amplitude $t$ and $\Theta\in[-\pi,\pi]$.

The hole levels on the QDs are described by the Hamiltonian
\begin{equation}\label{eq:H_h}
H_h=\sum_{D,\sigma}(\varepsilon_{hD}+\sigma\Delta_Z) h_{D\sigma}^\dagger h_{D\sigma} + U_{hD} \hat{n}_{hD\uparrow} \hat{n}_{hD\downarrow},
\end{equation}
where $h_{D\sigma}^{(\dagger)}$ annihilates (creates) a hole on QD $D$ with spin $\sigma=\uparrow,\downarrow$ and energy $\varepsilon_{hD}+\sigma\Delta_Z$, $U_{hD}$ is the charging energy and $\hat{n}_{hD\sigma}=h_{D\sigma}^\dagger h_{D\sigma}$ is the occupation number operator. We consider heavy holes with the $z$ component of the total angular momentum $j_{z,h\sigma}=\pm3\hbar/2$ \footnote{In small QDs, the confinement and geometry lead to the splitting of the heavy hole and light hole states, with the heavy hole states having the lowest energy. The light hole states are always empty and inaccessible to electron-hole recombination. Therefore, it is a good approximation to consider only the heavy hole states. For a review on this topic, see Ref. \cite{Lodahl2015}.}. Since the holes are energetically separated from the superconducting ground state by a large energy gap, we neglect a direct coherent coupling between the holes and the MBSs. Thus, the hole eigenstates are simply given by the occupation number states $\ket{\psi}_h=\ket{\hat{n}_{hA\sigma},\hat{n}_{hB\sigma}}$.

The hole reservoirs and their couplings to the holes on the QDs are given by
\begin{align}\label{eq:hole_reservoir}
&H_{\rm res}=\sum_{D,q,\sigma} \varepsilon_{Dq\sigma} h_{Dq\sigma}^\dagger h_{Dq\sigma},\\
&V_{h}=\sum_{D,q,\sigma} V_{Dq\sigma} h_{Dq\sigma}^\dagger h_{D\sigma} + \rm H.c.,\label{eq:hole_coupling}
\end{align}
respectively, where $h_{Dq\sigma}$ annihilates a hole in reservoir $D$ with spin $\sigma$ and energy $\varepsilon_{Dq\sigma}$. The hole refilling rates are given by $\Gamma_{hD\sigma}(\varepsilon)=(2\pi/\hbar) \sum_q |V_{Dq\sigma}|^2 \delta(\varepsilon-\varepsilon_{Dq\sigma})$, which we consider to be energy independent and that $\Gamma_{hD\sigma}=\Gamma_h$. Additionally, we assume that all hole states lie below $\mu_h$ such that holes are refilled on the QDs but cannot tunnel back to the reservoirs, see also Ref. \cite{Bittermann2022}.

The photon reservoirs are described by the Hamiltonian
\begin{align}\label{eq:H_ph}
H_{\rm ph}=\sum_{k,D,P} \hbar\omega_k a_{kDP}^\dagger a_{kDP},
\end{align}
where $a_{kDP}^{(\dagger)}$ annihilates (creates) a photon with wave number $k$ from QD $D$, polarization $P=L,R$, and energy $\hbar\omega_k$. The photons are circularly polarized with the $z$ component of the total angular momentum $j_{z,{\rm ph}}=\mp \hbar$ corresponding to left $(L)$ and right $(R)$ circular polarization, respectively.

Photons are emitted via electron-hole recombination,
\begin{align}\label{eq:V_rec}
V_{\rm rec}=g \sum_{k,D,\zeta} d_{D\zeta} h_{D\bar{\zeta}} a_{kD\zeta}^\dagger + {\rm H.c.},
\end{align}
with the light-matter-interaction energy $g$ and where we use $\zeta$ for the electron and hole spins and the photon polarization and identify $\sigma=\uparrow$ with $P=L$ and $\sigma=\downarrow$ with $P=R$. 

The Hamiltonian satisfies optical selection rules and holds for photons emitted in wire direction, where the total angular momentum commutes with the Hamiltonian, see for instance Refs. \cite{Rochat2000,Lodahl2015}. The conservation of the angular momentum, $j_{z,e\uparrow}+j_{z,h\downarrow}=-\hbar$ ($L$ photons) and $j_{z,e\downarrow}+j_{z,h\uparrow}=+\hbar$ ($R$ photons), leads to selection rules where $\uparrow$($\downarrow$) electrons recombine with $\downarrow$($\uparrow$) holes to $L(R)$ photons, and thus, allowing us to establish a correspondence between the polarization of the emitted photon and the electron spin. Photons emitted in other directions would have a different polarization and we would have to adjust $V_{\rm rec}$ accordingly \cite{Cerletti2005,Gywat2002}. We can enhance the number of photons emitted in wire direction by integrating the nanowire in a photonic waveguide \cite{Zadeh2016}.

Note that we focus on electron-hole recombination processes rather than excitonic effects caused by electron-hole interactions. The latter can be reduced by applying an electric field that spatially separates the electron and hole wave functions \cite{Polland1985,Plentz1997}. Thereby, also the photon emission rate can be decreased, ensuring that the dynamics of the setup is governed by the coherent dynamics of electrons in the QD-MBSs system ($|t_{D\sigma}/\hbar|\gg\Gamma_\textrm{ph}$).

Note that using a nanowire has a practical importance on both the Majorana and optics sides. Indeed, previous experiments have successfully demonstrated the coherent tunnel-coupling of QDs to Rashba-nanowires in proximity to an $s$-wave superconductor \cite{Deng2016,Deng2018} and the combination of superconductors with semiconductor optics \cite{Sasakura2011,Panna2018,Yang2021} as well as the embedment of optically active QDs within $pn$ junctions \cite{Minot2007,Versteegh2014}. For these reasons, we believe that by combining these elements, our setup is experimentally feasible.

\section{Master equation and counting statistics}\label{sec:master}

To investigate the dynamics of the system and its transport properties, we use a Markovian master equation approach \cite{Flindt2004,Kaiser2007,Emary2007,Sanchez2008,Dominguez2010}. To this aim, we trace out the photons and the hole reservoirs from the total Hamiltonian, allowing us to study the dynamics of the central part consisting of the MBSs and the QDs, $H_S=H_e+H_h$. Using standard approximations one can arrive to 
\begin{align}\label{eq:mastereq}
\begin{split}
\partial_t\rho^{\psi\psi}_{S}(\chi_{D\zeta},t)&= \sum_{\psi'} 
\left[
-W^{D\zeta}_{\psi'\psi}\rho^{\psi\psi}_{S}(\chi_{D\zeta},t)\right.\\ 
&\left.+ e^{i\chi_{D\zeta}} W^{D\zeta}_{\psi\psi'} \rho^{\psi'\psi'}_{S}(\chi_{D\zeta},t)\right],
\end{split}
\end{align}
where the counting field $\chi_{D\zeta}$ counts photons emitted from QD $D$ with polarization $\zeta$.
Here, the diagonal elements of the density matrix in the system state $\rho_S^{\psi\psi}=\bra{\psi}\rho_S(t)\ket{\psi}$ describe the occupation probability of state $\ket{\psi}=\ket{\psi}_e\times\ket{\psi}_h$ at time $t$. Moreover, $W^{D\zeta}_{\psi\psi'}$ are the electron-hole recombination rates 
\begin{align}
W^{D\zeta}_{\psi\psi'}=\Gamma_{\rm ph}
|\bra{\psi}d_{D\zeta}h_{D\bar{\zeta}}\ket{\psi'}|^2,
\end{align}
with the photon rate $\Gamma_{\rm ph}=\frac{2\pi}{\hbar}\nu_{\rm ph}|g|^2$ and the photon density $\nu_{\rm ph}$. Here, we consider no photons in the initial state, thus, photons can only be emitted, see further details of the calculations in App.~\ref{app:master_equation}.

We supplement the master equation by the additional rates,
\begin{align}
\mathcal L_{\rm qp}^{D\sigma}[\rho_S]=\Gamma_{\rm qp}(d_{D\sigma}^\dagger\rho_S d_{D\sigma}-\frac{1}{2}\{\rho_S, d_{D\sigma} d_{D\sigma}^\dagger \}),
\end{align}
\begin{align}
\mathcal L_{h}^{D\sigma}[\rho_S]=\Gamma_{h}(h_{D\sigma}^\dagger\rho_S h_{D\sigma}-\frac{1}{2}\{\rho_S, h_{D\sigma} h_{D\sigma}^\dagger \}).
\end{align}
Here, the rate $\Gamma_{\rm qp}$ describes a process, where uncorrelated quasiparticles  occupy the QDs, and the rate $\Gamma_h$, that can be derived from Eqs.~\eqref{eq:hole_reservoir} and~\eqref{eq:hole_coupling} \cite{Blum2010}, refills holes with spin $\sigma$ on QD $D$ from the normal reservoir. We assume a large hole refilling rate $\Gamma_{h}\gg \Gamma_{\rm ph}$ and $|t_{D\sigma}/\hbar|\gg \Gamma_{\rm ph}$, such that the dynamics of the system is fully governed by the processes in the QD-MBSs system \footnote{Note that in the limit of a slow hole refilling ($\Gamma_{h}\ll\Gamma_{\rm ph}\ll|t_{D\sigma}/\hbar|$), the total cross-noise would be negative, since the dynamics of the system would mostly be governed by the hole subsystem which acts like a single resonant level without superconducting correlations.}. We include these processes in the master equation by taking the expectation value in the system state and adding them to the right-hand side of Eq.~\eqref{eq:mastereq}. 
\begin{figure}
	\includegraphics[width=\columnwidth]{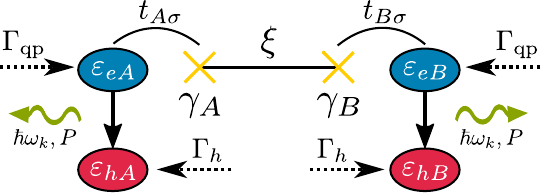}
	\caption{Sketch of the dynamics in the model. We show the MBSs $\gamma_A$ and $\gamma_B$ with splitting $\xi$ coupled to electrons on the QDs with energies $\varepsilon_{eD}$ with amplitudes $t_{D\sigma}$. The holes on the QDs with energies $\varepsilon_{hD}$ get refilled by the hole refilling rate $\Gamma_h$. Via electron-hole recombination photons are emitted with energy $\hbar\omega_k$ and polarization $P$. Uncorrelated quasiparticles can occupy the electronic QD states with rate $\Gamma_{\rm qp}$.}
	\label{fig:setup_scheme}
\end{figure}

The time evolution of the occupations is calculated from the master equation
\begin{align}
\partial_t\rho_{S}(\chi_{D\zeta},t)=\mathcal{L}(\chi_{D\zeta})\rho_{S}(\chi_{D\zeta},t),
\end{align}
where $\rho_{S}(\chi_{D\zeta},t)$ is a vector whose entries contain the occupation of each state $\ket{\psi}$, i.e.~$\rho^{\psi\psi}_{S}(\chi_{D\zeta},t)$. In this form, the Liouvillian $\mathcal{L}(\chi_{D\zeta})$ is a matrix containing all rates connecting different states, see a schematic picture of the setup in Fig.~\ref{fig:setup_scheme}.

The stationary state $\rho_{\rm stat}$ is obtained by solving $\partial_t\rho_S(\chi_{D\zeta}=0,t)=0$. This is equivalent to find the eigenvector $\ketvec{\phi_0}$ of the Liouvillian $\mathcal{L}=\mathcal{L}(\chi_{D\zeta}=0)$ with zero eigenvalue $\lambda_0=0$. The components of this column vector $\rho_{\rm stat}$ contain the stationary occupations of each state $\ket{\psi}$. From orthonormality $\braketvec{\phi_0|\phi_0}=1$, it follows that $\bravec{\phi_0}$ is a row vector with entries of 1, so that applying $\bravec{\phi_0}$ from the left corresponds to taking the trace.

Correlations of emission events are given by the noise power spectrum \cite{Blanter2000}
\begin{align}\label{eq:noisepower}
S_{D\zeta,D'\zeta'}(\omega)=\frac{1}{2} \int_{-\infty}^\infty dt e^{i\omega t} \braket{\{\delta I_{D\zeta} (t) ,\delta I_{D'\zeta'}(0)\}},
\end{align}
with $D,D'=A,B$ and the current fluctuations $\delta I_{D\zeta}(t)=I_{D\zeta}(t)-\braket{I_{D\zeta}(t)}$ of photons with spin $\zeta$ emitted from QD $D$. Here, curly brackets denote the anticommutator and $\omega$ and $t$ frequency and time, respectively. In Eq.~\eqref{eq:noisepower}, $\langle ... \rangle=\rm{Tr}[\rho_0 ...]$ with $\rho_0$ the equilibrium density matrix of the system and bath. The current operator $I_{D\zeta}(t)={\dot N_{D\zeta}}$ with $N_{D\zeta}=\sum_ka_{kD\zeta}^{\dagger}a_{kD\zeta}$ the number operator for photons associated with emission from QD $D$. From Eq.~\eqref{eq:noisepower}, we can obtain the autocorrelation function $S_{D\zeta,D\zeta'}$ for correlations in a single QD $D=D'=A/B$ or the cross-correlation function $S_{D\zeta,D'\zeta'}$ with $D=A$ and $D'=B$. In what follows, we restrict the calculations to the zero-frequency noise ($\omega=0$) of the cross-correlations. Therefore, we simplify the notation removing the $D,D'$ labels and the $\omega$ dependence. In this way, we can express the cross-correlation via the current superoperators $\mathcal{J}_{D\zeta}$ as \cite{Flindt2004}
\begin{align}\label{eq:crossnoise}
&S_{\zeta\zeta'}=-\bravec{\phi_0}(\mathcal{J}_{A\zeta}\mathcal{R}\mathcal{J}_{B\zeta'}+\mathcal{J}_{B\zeta'}\mathcal{R}\mathcal{J}_{A\zeta})\ketvec{\phi_0},
\end{align}
which describes the correlation between two photons emitted from QD $A$ with spin $\zeta$ and QD $B$ with spin $\zeta'$. Here, we have introduced the jump superoperators
\begin{align}
\mathcal{J}_{D\zeta}=-i \partial_{\chi_{D\zeta}}\mathcal{L}(\chi_{D\zeta})\Big|_{\chi_{D\zeta}=0},
\end{align}
which describe the process of photon emission from QD $D$ with spin $\zeta$ \footnote{We could also define a current cross-noise by counting holes that left a QD, since holes can only leave the QD via photon emission. We would just have to add the elementary charge $e$ to Eq.~\eqref{eq:crossnoise} accordingly.}.
Moreover, we have used the projectors $\mathcal{P}_0=\ketvec{\phi_0}\bravec{\phi_0}$ and $\mathcal{Q}=1-\mathcal{P}_0$, and defined the pseudoinverse of the Liouvillian with $\mathcal{R}=\mathcal{Q}\mathcal{L}^{-1}\mathcal{Q}$.

\begin{figure}[t]
	\includegraphics[width=\columnwidth]{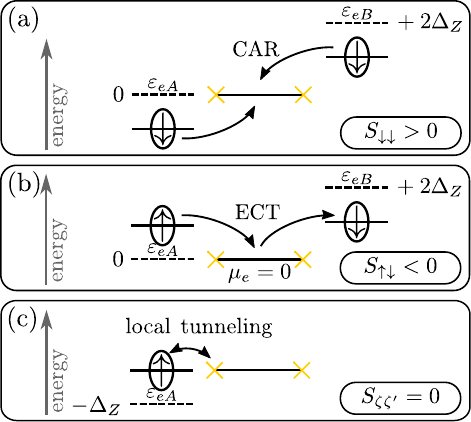}
	\caption{Types of resonances in the QD-MBSs system. We show the electronic QD levels relative to the superconducting condensate ($\mu_e=0$). For $\varepsilon_{eA}=0$ and $\varepsilon_{eB}=2\Delta_Z$, CAR processes lead to a positive cross-correlation $S_{\down\down}>0$ (a), whereas ECT leads to a negative contribution of $S_{\uparrow\downarrow}$ (b). For $\varepsilon_{eA}=-\Delta_Z$, we show the mechanism of local tunnel processes for $\uparrow$ electrons on QD $A$ (c), which does not contribute to the cross-correlations.}
	\label{fig:resonances}
\end{figure}

If emission from the two QDs is correlated, we have $S_{
\zeta\zeta'}\neq 0$, otherwise $S_{\zeta\zeta'}=0$. In our model, the only coherent coupling between the two QDs mediating such correlations proceeds via the finite hybridization energy $\xi$. The size and sign of $S_{\zeta\zeta'}$ further depends crucially on the QD energies as well as on $\zeta$ and $\zeta'$. As we will discuss in detail below, the coherent coupling of two electron spins (one in each QD) via the superconducting condensate is reminiscent of crossed Andreev reflection (CAR), i.e. the splitting of a Cooper pair via the two QDs. Subsequent correlated emission of two photons with spins $\zeta$ and $\zeta'$ from the two QDs leads to $S_{\zeta\zeta'}>0$, see Fig.~\ref{fig:resonances}(a). On the contrary, if only one electron is shared between the two QDs, the coherent tunneling between the two QDs via the two MBSs is reminiscent of elastic cotunneling (ECT) and leads to negative cross-noise $S_{\zeta\zeta'}<0$ as only one photon is emitted from a QD, whereas the other QD will not emit a photon at the same time, see Fig.~\ref{fig:resonances}(b).

In practice, photons emitted from one of the QDs can be detected on both sides of the wire. The distinction of the emission from the two QDs, necessary for measuring cross-correlations, could be ensured by having different emission energy ranges for the two QDs by shifting the hole states on one side with a gate voltage. In that case the location of emission is correlated with the energy of the photons that can be readily measured by a photo detector.

\begin{figure*}[t]
	\includegraphics[width=\textwidth]{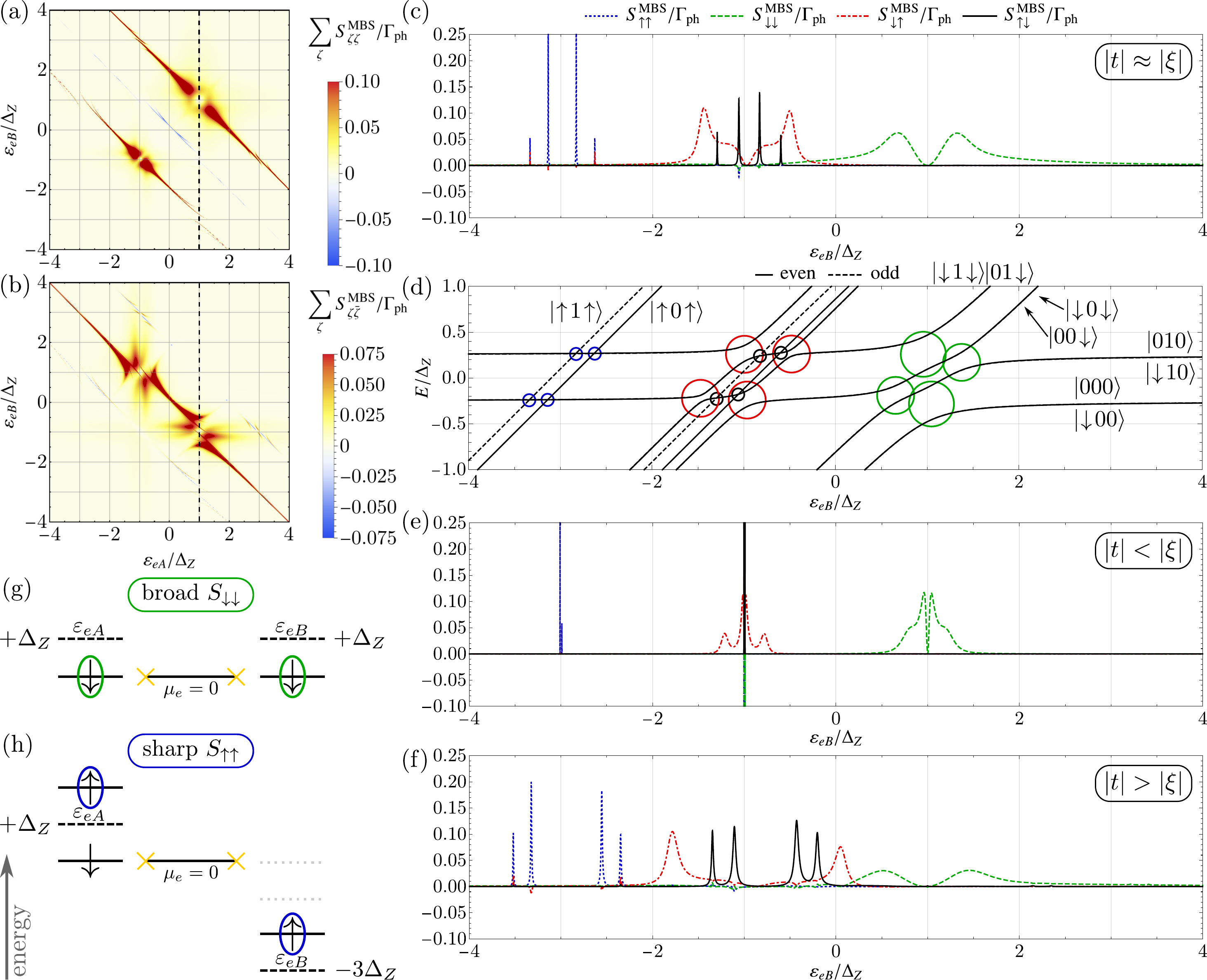}
	\caption{Cross-correlations spectroscopy for MBSs. We show $\sum_\zeta S_{\zeta\zeta}$ in (a) and $\sum_\zeta S_{\zeta\bar\zeta}$ in (b) for parallel and anti-parallel spin configurations, respectively, as a function of the QD energies $\varepsilon_{eA}$ and $\varepsilon_{eB}$ in the intermediate tunneling regime ($|t|\approx|\xi|$). The dashed lines correspond to the line cut in (c), where $S_{\zeta\zeta'}$ is plotted along $\varepsilon_{eA}=\Delta_Z$. We give the energy spectrum at $\varepsilon_{eA}=\Delta_Z$ as a function of $\varepsilon_{eB}$ in (d), where even- (solid lines) and odd-parity eigenstates (dashed lines) show anticrossings indicated by circles with colors that correspond to the peaks of $S_{\zeta\zeta'}$ in panel (c). We show two additional line cuts for the small ($|t|<|\xi|$) and large ($|t|>|\xi|$) tunneling regimes in panel (e) and (f), respectively. The parameters are $\xi=0.2\Delta_Z$, $\Theta=0.75\pi$, and $\Gamma_{\rm qp}=0.$ The tunneling amplitude is $t=0.25\Delta_Z$, except in (e) we use $t=0.05\Delta_Z$ and in (f) $t=0.5\Delta_Z$. Furthermore, we give schematic figures for the broad $S_{\down\down}$ and the sharp $S_{\up\up}$ resonances depending on the energy level position $\varepsilon_{eB}$ in panels (g) and (h).}
	\label{fig:MBS}
\end{figure*}
\section{Cross-correlations spectroscopy}\label{sec:correlations}

In this section, we first investigate the cross-correlations spectroscopy for a $pnp$ junction made of a 1D TSC coupled to two QDs.
We demonstrate that our detection scheme can probe (i) the presence of nonlocal triplet superconducting correlations, which is a necessary condition for the emergence of MBSs, and (ii) establish a correspondence between the number of resonance peaks and the number of MBSs coupled to the QDs. Second, we substitute the MBSs with trivial qMBSs comprising four coupled MBSs and compare the results of the cross-correlations spectroscopy.

Due to the different nature of CAR and ECT processes, their resonances occur under different conditions: CAR processes require that the energies of the electrons on the QDs are compensated relatively to the superconducting condensate ($\mu_e=0$) \cite{Recher2001}, i.e.~$\varepsilon_{eA}+\sigma\Delta_Z=-(\varepsilon_{eB}+\sigma'\Delta_Z)$, leading to positive cross-correlations and the emission of a highly-correlated photon pair, whereas, ECT processes require an alignment of the electron levels $\varepsilon_{eA}+\sigma\Delta_Z=\varepsilon_{eB}+\sigma'\Delta_Z$, leading to negative cross-correlations.

Aside from CAR and ECT, local emission processes can reduce the cross-correlations magnitude as they reduce the number of nonlocal emission events. However, these processes are resonant when the QD spin level is aligned with the superconductor $\varepsilon_{eD}+\sigma\Delta_Z=0$ \cite{Bittermann2022}, see Fig.~\ref{fig:resonances}(c).

In summary, we expect CAR which leads to positive cross-correlations along the anti-diagonal $\varepsilon_{eA}=-\varepsilon_{eB}$ for $S_{\zeta\bar{\zeta}}$ and along $\varepsilon_{eA}=-\varepsilon_{eB}\pm2\Delta_Z$ for $S_{\zeta\zeta}$. On the other hand, ECT gives rise to negative cross-correlation resonances along the diagonal $\varepsilon_{eA}=\varepsilon_{eB}$ for $S_{\zeta\zeta}$ and $\varepsilon_{eA}=\varepsilon_{eB}\pm2\Delta_Z$ for $S_{\zeta\bar{\zeta}}$. Besides, these resonances can be reduced along $\varepsilon_{eB}=\pm\Delta_Z$ and $\varepsilon_{eA}=\pm\Delta_Z$ due to local tunneling processes. Note that while CAR and ECT processes leave the parity unchanged, local emission processes change the parity.

\subsection{Majorana bound states}\label{sec:MBS}

We first explore the nonlocal superconducting correlations by exploiting the spin texture in the analysis of the cross-correlations spectroscopy. Thus, we split $S_{\zeta\zeta'}$ into its parallel $S_{\zeta\zeta}$ and anti-parallel $S_{\zeta\bar{\zeta}}$ spin components and represent them as a function of the QD energies $\varepsilon_{eA}$ and $\varepsilon_{eB}$, see Fig.~\ref{fig:MBS}.
In panels~(a) and (b), we can observe three positive resonance cross-correlation lines for the intermediate tunneling regime ($|t|\approx|\xi|$) that fulfill the condition for CAR, see Fig.~\ref{fig:resonances}(a). In Fig.~\ref{fig:MBS}(a), we observe two resonance lines extending along $\varepsilon_{eA}=-\varepsilon_{eB}\pm2\Delta_Z$, which correspond to the triplet QD states with spin $m_s=\mp1$. In turn, the resonance line along $\varepsilon_{eA}=-\varepsilon_{eB}$, observed in panel (b), emerges from singlet and triplet states with spin $m_s=0$. As we mentioned above, triplet correlations serve as a necessary condition for the existence of MBSs in $p$-wave superconductors. However, we will observe later that trivial qMBSs lead to comparable resonance lines, differing primarily in the number of resonance peaks along a line cut, where one QD energy remains constant.

In order to find more specific features in the spectroscopy of $S_{\zeta\zeta'}$ that allow us to distinguish between MBSs and qMBSs, we analyze $S_{\zeta\zeta'}$ along $\varepsilon_{eA}=\Delta_Z$ in Fig.~\ref{fig:MBS}(c), where the $\down$ electron on QD~$A$ is at the Fermi energy $\mu_e$ of the SC. Along this line, we observe two types of CAR resonances, broad and sharp, corresponding to strong and weak coupling to the superconductor, respectively, which is effectively determined by the arrangement of the QD energy levels: broad (sharp) resonances involve both (none) of the QD energy levels close to resonance with the superconductor ($\mu_e=0$). For the linecut along $\varepsilon_{eA}=\Delta_Z$, the $\down$ electron on QD~$A$ is in resonance with the superconductor, yielding broad resonances with $S_{\downarrow \zeta}$, see Fig.~\ref{fig:MBS}(g). Sharp resonances correspond to the opposite spin configuration $S_{\uparrow \zeta}$, see Fig.~\ref{fig:MBS}(h).

Broad and sharp resonances can be linked to the presence of an anticrossing on the energy spectrum, see circles in the same color in Fig.~\ref{fig:MBS}(d). In general, we observe up to $2^{\#\text{MBSs}}=4$ peaks for each component $S_{\zeta\zeta'}$ of the cross-correlations resulting from the hybridization of the QD levels and the two fermionic states that form the coupled MBSs.
Naturally, the number of resonances can be reduced due to the presence of degeneracies or the overlapping of peaks in larger tunneling regimes ($|t|\geq|\xi|$). Thus, having four peaks per cross-correlation component serves as an upper bound in the scenario when two MBSs are present in the system. Note also that since we choose a spin angle of $\Theta=0.75\pi$, the tunneling of $\down$ electrons is favored, thus $S_{\zeta\down}$ exhibits broader resonance peaks than $S_{\zeta\up}$.

To cover different scenarios,
we analyze $S_{\zeta\zeta'}$ for different tunneling amplitude strengths relative to the splitting energy $\xi$ of the MBSs.
For $|t|<|\xi|$, see Fig.~\ref{fig:MBS}(e), $\xi$ determines the energy splitting between eigenstates of even and odd parity given in the product basis introduced in Sec.~\ref{sec:model} (e.g. $\ket{0,0,0}$ and $\ket{0,1,0}$). Here, the sharp resonances are hardly visible, since they become very thin. Moreover, the broad resonances exhibit a four-peak structure. For the $S_{\down\down}$ component, the two outer peaks are located at $\varepsilon_{eB}=\Delta_Z\pm\xi$. They result from emission cycles that involve the most contributing states, 
\begin{align}\label{eq:emissioncycle1}
\ket{0,0,0}\leftrightarrow\ket{\down,1,0}\Leftrightarrow\ket{\down,0,\down}\xrightarrow{\text{2 ph}}\ket{0,0,0} ,
\end{align}
in the even-parity sector for $\varepsilon_{eB}=\Delta_Z-\xi$, whereas
\begin{align}\ket{0,1,0}\leftrightarrow\ket{\down,0,0}\Leftrightarrow\ket{\down,1,\down}\xrightarrow{\text{2 ph}}\ket{0,1,0},
\end{align}
in the odd-parity sector at $\varepsilon_{eB}=\Delta_Z+\xi$.
Here, the hybridization is denoted to be strong ($\Leftrightarrow$) or weak ($\leftrightarrow$). Then, the two inner peaks close to $\varepsilon_{eB}=\Delta_Z$ arise from both,
\begin{align}
\ket{0,0,0}\leftrightarrow\ket{\down,0,\down}\xrightarrow{\text{2 ph}}\ket{0,0,0},\\
\ket{0,1,0}\leftrightarrow\ket{\down,1,\down}\xrightarrow{\text{2 ph}}\ket{0,1,0},
\end{align}
even- and odd-parity states.
Remarkably, exactly at resonance ($\varepsilon_{eB}=\Delta_Z$), the cross-correlations become zero, since CAR and ECT processes compensate each other. This is because the eigenstates are equal superpositions, e.g. in the even parity,
\begin{align}
\ket{\down,1,0}\Leftrightarrow\ket{0,1,\down} \ (E>0) ,\\
\ket{0,0,0}\Leftrightarrow\ket{\down,0,\down}\xrightarrow{\text{2 ph}}\ket{0,0,0} \ (E<0),
\end{align}
which are equally connected via ECT or CAR, respectively, with the energy $E$ of eigenstates of $H_e$ in Eq.~\eqref{eq:H_e,fermion}. In addition, local emission becomes dominant due to the presence of degeneracies between even- and odd-parity states involved in emission cycles for single photon emission. Thus, at $\varepsilon_{eB}=\Delta_Z$ for $E>0$,
hybridization enables the emission cycles
\begin{align}
\ket{\down,1,\down}\xrightarrow{\text{1 ph}}\ket{0,1,\down}\xrightarrow{\text{1 ph}}\ket{0,1,0}\Leftrightarrow\ket{\down,1,\down},\\
\ket{\down,1,\down}\xrightarrow{\text{1 ph}}\ket{\down,1,0}\xrightarrow{\text{1 ph}}\ket{0,1,0}\Leftrightarrow\ket{\down,1,\down}. \label{eq:emissioncycle2}
\end{align}

Furthermore, when we only take one spin species into account, which is the case, for instance, when considering a Majorana angle of $\Theta=0,\pi$, or in the limit of $|t_{D\sigma}|\ll|\Delta_Z|$, we can reduce the system to a spinless model. In the limit of $|t_{D\sigma}|/|\varepsilon_{eD\sigma}-\xi|\ll 1$, we can further provide an analytical expression for the cross-correlations, by means of a Schrieffer-Wolff transformation,
\begin{align}\label{eq:S_ana}
&\tilde{S}_{\sigma\sigma}=\Gamma_{\rm ph}\frac{2 \tilde\Delta_p^2\big[2 \tilde\Delta_p^2 +(\varepsilon_{eA\sigma}+\varepsilon_{eB\sigma})^2\big]}{\big[4 \tilde\Delta_p^2+(\varepsilon_{eA\sigma}+\varepsilon_{eB\sigma})^2\big]^2},\\
&\tilde\Delta_p=i t_{A\sigma} t_{B\sigma} \left(\frac{\xi}{\varepsilon_{eA\sigma}^2-\xi^2}+\frac{\xi}{\varepsilon_{eB\sigma}^2-\xi^2}\right),
\end{align}
with $\varepsilon_{eD\sigma}=\varepsilon_{eD}+\sigma\Delta_Z$. The cross-correlations $\tilde{S}_{\sigma\sigma}$ exhibit similar features to those of $S_{\down\down}$ in the full model, as for the latter, the photon emission process predominantly involves a single spin species, see Eqs. \eqref{eq:emissioncycle1}-\eqref{eq:emissioncycle2}. Further details of the spinless model are presented in App.~\ref{app:spinless}.

In the limit of larger tunneling amplitudes ($|t|\geq|\xi|$), the cross-correlation peaks broaden and separate, see Fig.~\ref{fig:MBS}(c) and (f). The four-peak structure of the broad resonances evolves into two peaks, as a result of the overlapping of peaks caused by a larger hybridization.
However, the sharp resonances now exhibit four peaks, once again signifying the maximum number of peaks for two MBSs. The emergence of the two peaks in $S_{\uparrow\uparrow}$ for energies $\varepsilon_{eB}\gtrsim-3\Delta_Z$, for instance, can be attributed to the emission cycles,
\begin{align}
&\ket{0,1,0}\Leftrightarrow\ket{\up,1,\up}\xrightarrow{\text{2 ph}}\ket{0,1,0},\\
&\ket{\down,1,0}\Leftrightarrow\ket{\up,0,\up}\xrightarrow{\text{2 ph}}\ket{0,0,0}\leftrightarrow\ket{\down,1,0}.
\end{align}
Here, the first emission cycle directly connects the CAR coupled states, whereas the second cycle additionally includes a spin-flip process. Thus, the peak closer to $\varepsilon_{eB}=-3\Delta_Z$ is larger and broader. The separation between the two peaks is approximately given by $\xi$, since the corresponding anticrossings have a distance of $\Delta\varepsilon_{eB}\approx\xi$.

Comparing the large ($|t|\geq|\xi|$) and small ($|t|<|\xi|$) tunneling regimes, we observe that the latter constitutes the most difficult regime to resolve all four sharp peaks. We can understand this observation by differentiating the roles played by $\xi$ and $t$. While $\xi$ represents an energy difference between the unoccupied and occupied nonlocal fermion level, $t$ is the tunneling amplitude between QD electrons and the nonlocal fermion.
Thus, cross-correlation peaks, emerging at resonance (when the CAR or ECT condition is fulfilled), decrease as $|t|^2/|\xi|$ when detuning the QD levels from resonance. This occurs because $\xi$ increases the energy cost for virtually occupying the nonlocal fermion. As a result, discerning resonances is more challenging (easier) when $|t|<|\xi| ~(|t|\geq|\xi|)$, because it gives rise to sharper (broader) peaks.

In reality, the finite size of the TSC also allows for nonlocal tunnel couplings between QDs and MBSs on opposite sides \cite{Prada2017,Schuray2017,Clarke2017}. For the discussed regime of overlapping MBSs, experimental data from Ref.~\cite{Deng2018} demonstrates, that the ratio between the nonlocal and local tunneling amplitudes is small, yielding no qualitative differences compared to the case without nonlocal couplings. Note that a higher ratio could potentially lead to the emergence of additional resonance peaks, which could be erroneously associated with trivial states, see more details in App.~\ref{app:MBSnl}.

\begin{figure}[t]
	\includegraphics[width=\columnwidth]{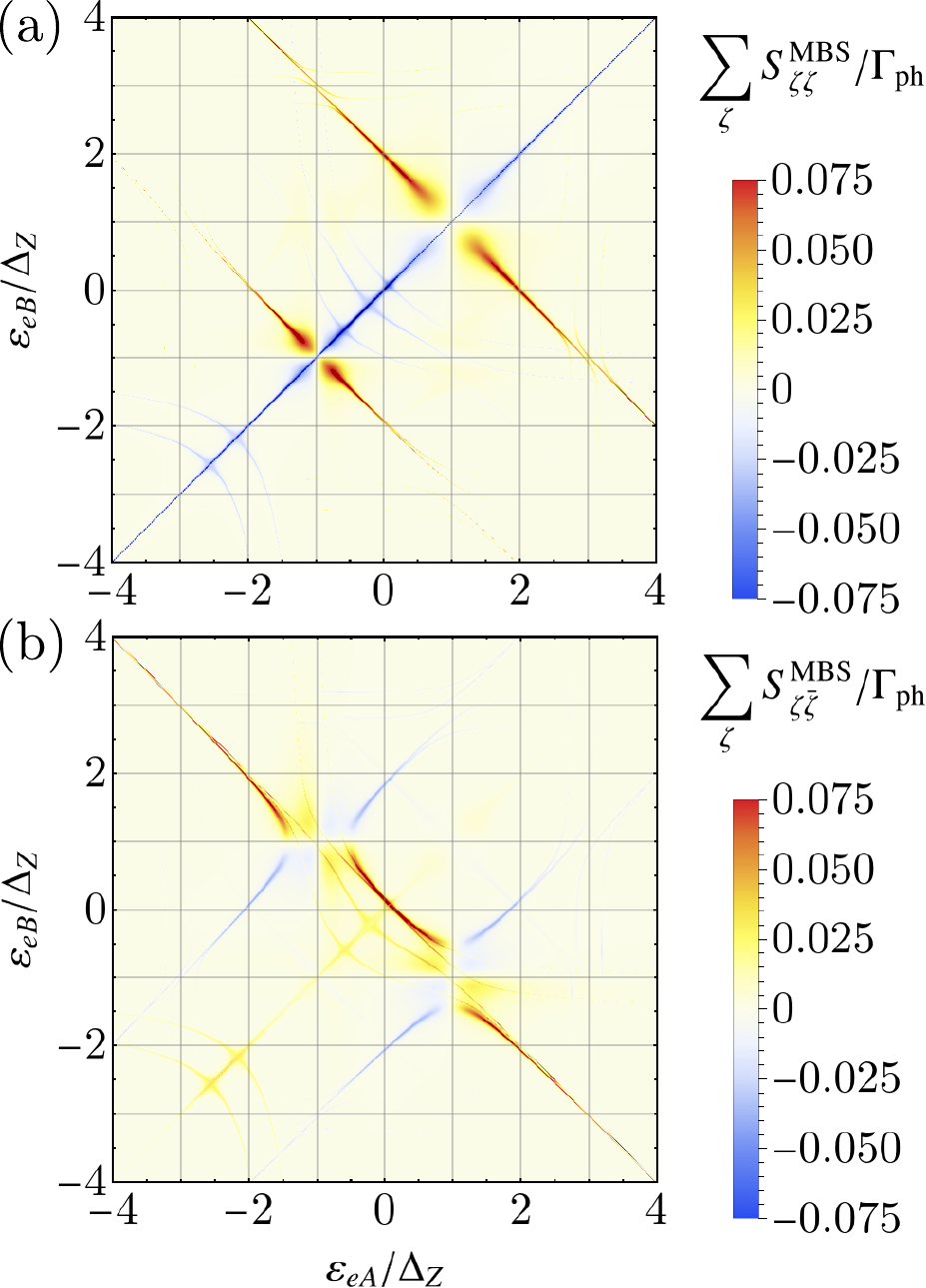}
	\caption{Cross-correlations spectroscopy for MBSs with finite $\Gamma_{\rm qp}$. We show $\sum_\zeta S_{\zeta\zeta}$ in (a) and $\sum_\zeta S_{\zeta\bar\zeta}$ in (b) for parallel and anti-parallel spin configuration, respectively, as a function of the QD energies $\varepsilon_{eA}$ and $\varepsilon_{eB}$. The parameters are $t=0.25\Delta_Z$, $\xi=0.2\Delta_Z$, $\Theta=0.75\pi$, $\Gamma_{\rm qp}=\Gamma_{\rm ph}/2\ll|t_{D\sigma}/\hbar|$.}
	\label{fig:MBS_cont}
\end{figure}

Now we perturb the system to check the robustness of the nonlocal cross-correlations spectroscopy by adding a constant rate of uncorrelated quasiparticles. 
The presence of additional uncorrelated quasiparticles modifies the cross-correlation signal as local emission events are not limited to the resonance condition between the QD levels and the condensate, but they are present for all QD energies $\varepsilon_{eD}$. Furthermore, ECT processes are enhanced, since uncorrelated quasiparticles can tunnel between the QDs via the MBSs when the energy levels are aligned, i.e. $\varepsilon_{eA}+\sigma\Delta_Z=\varepsilon_{eB}+\sigma'\Delta_Z$. Note that the constant rate changes the parity of the system and that it needs to be small ($\Gamma_{\rm qp}\ll|t_{D\sigma}/\hbar|$), such that the system stays coherent.
\begin{figure*}[t]
	\includegraphics[width=\textwidth]{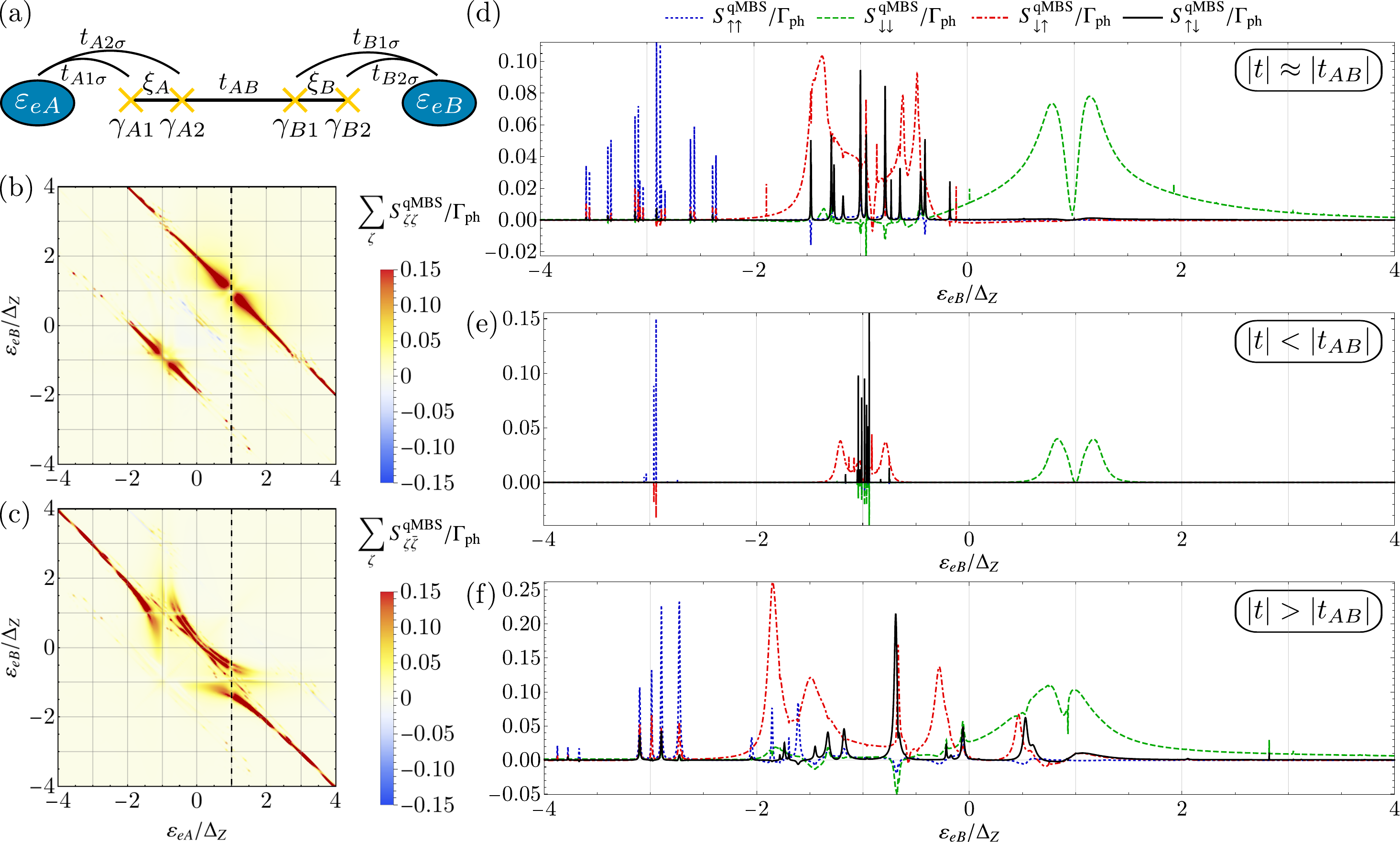}
	\caption{Cross-correlations spectroscopy for qMBSs. We give a schematic figure of the qMBS model in panel (a). We show $\sum_\zeta S_{\zeta\zeta}$ in (b) and $\sum_\zeta S_{\zeta\bar\zeta}$ in (c) for parallel and anti-parallel spin configurations, respectively, as a function of the QD energies $\varepsilon_{eA}$ and $\varepsilon_{eB}$ in the intermediate tunneling regime ($|t|\approx|t_{AB}|$). The dashed lines correspond to the line cut in (d), where $S_{\zeta\zeta'}$ is plotted along $\varepsilon_{eA}=\Delta_Z$. We show two additional line cuts for the small ($|t|<|t_{AB}|$) and large ($|t|>|t_{AB}|$) tunneling regimes in panel (e) and (f), respectively. The parameters are $\xi_A=\xi_B=0$, $t_{AB}=0.1\Delta_Z$, $\Theta_A=\Theta_B=0.75\pi$, and $\Gamma_{\rm qp}=0.$ The tunneling amplitude is $t=0.25\Delta_Z$, except in (e) we use $t=0.05\Delta_Z$ and in (f) $t=0.5\Delta_Z$.}
	\label{fig:qMBS}
\end{figure*}

We show in Fig.~\ref{fig:MBS_cont} the cross-correlations spectroscopy in the presence of uncorrelated quasiparticles. We can observe that positive cross-correlations are still present, although they become reduced at all energies $(\varepsilon_{eA},\varepsilon_{eB})$ due to the enhanced local emission. By comparing Figs.~\ref{fig:MBS} and~\ref{fig:MBS_cont}, additional negative resonances (blue) appear along the diagonal that where absent in the case of $\Gamma_{\rm qp}=0$, resulting from ECT. For parallel spins at $\varepsilon_{eA}=\varepsilon_{eB}$, a spin can tunnel from one QD to the opposite one, see Fig.~\ref{fig:MBS_cont}(a), whereas for anti-parallel spins at $\varepsilon_{eA}=\varepsilon_{eB}\pm2\Delta_Z$, a spin needs to flip while tunneling through the system, see Fig.~\ref{fig:MBS_cont}(b).

Note that adding uncorrelated quasiparticles directly on the superconductor \cite{Rainis2012} has no significant effect on the cross-correlation signal, if the rate is smaller or comparable to $\Gamma_\textrm{ph}$, as the emission of photon pairs occurs in both parity sectors. Thus, the proposed effects of our system exhibit a certain degree of immunity against various kinds of quasiparticle poisoning.

In summary, the cross-correlations spectroscopy reveals the presence of finite triplet correlations, which is a necessary condition for the emergence of MBSs. Furthermore, we observe up to $2^{\#\text{MBSs}}=4$ peaks for each component $S_{\zeta\zeta'}$ as an upper bound, establishing a correspondence between the number of MBSs and the number of peaks in the spectroscopy.
Additionally, the cross-correlations remain robust even in the presence of uncorrelated quasiparticles, provided that the poisoning rate fulfills $\Gamma_{\rm qp}\ll|t_{D\sigma}/\hbar|$.

\subsection{Quasi-Majorana bound states}\label{sec:qMBSs}

We now add two extra MBSs to the system studied above to mimic a scenario where nontopological qMBS are present in the system, see Fig.~\ref{fig:qMBS}(a). We thus introduce the Majorana operators $\gamma_{D1}=c_D^\dagger+c_D$ and $\gamma_{D2}=i(c_D^\dagger-c_D)$ that comprise a complex fermion $c_D^\dagger$ coupled to QD $D=A,B$.
To compare our previous results for the MBSs case with those of the qMBSs, we replace Eq.~\eqref{eq:H_e,fermion} by the model that describes the qMBSs coupled to the QDs in the basis of the two complex fermions,
\begin{align}\label{eq:qMBS}
\begin{split}
H_{e}^{\rm qMBS}=&\sum_{D,\sigma}(\varepsilon_{eD}+\sigma\Delta_Z) d_{D\sigma}^\dagger d_{D\sigma}+\sum_D\xi_D \left(\hat{n}_{cD} -\frac{1}{2}\right)\\
&+t_{AB}(c_Ac_B-c_A^\dagger c_B) +\rm H.c.\\
&+\sum_{D,\sigma} (t_{D1\sigma}+it_{D2\sigma})d_{D\sigma} c_D^\dagger \\
&+(t_{D1\sigma}-it_{D2\sigma})d_{D\sigma} c_D +\rm H.c..
\end{split}
\end{align}
Here, the first term describes the two QDs with energy $\varepsilon_{eD}+\sigma\Delta_Z$ and a Zeeman splitting $\Delta_Z$. The second term is the coupling $\xi_D$ of two MBSs $\gamma_{Di}$, $i=1,2$, on the same side $D=A,B$ with occupation number operator $\hat{n}_{cD}=c_D^\dagger c_D$, where $c_D^{(\dagger)}$ annihilates (creates) a complex fermion on side $D=A,B$. The third term accounts for the coupling $t_{AB}$ between the two complex fermions on different sides that is necessary to obtain finite cross-correlations. It results from the coupling
of the inner MBSs $\gamma_{A2}$ and $\gamma_{B1}$. The last term describes the tunneling between the MBSs and the QDs with the spin-dependent tunneling amplitudes,
\begin{align}
t_{D1\up}=-it_{D2\up} = t \cos(\Theta_D/2),\\
t_{D1\down}=+it_{D2\down} = t \sin(\Theta_D/2),
\end{align}
which we can parametrize by the angle $\Theta_D$ for the complex fermion on side $D$.

We diagonalize $H_e^{\rm qMBS}$ in Eq.~\eqref{eq:qMBS} in the basis $\ket{n_{dA\sigma},n_{cA},n_{cB},n_{dB\sigma}}$ and calculate the cross-correlations. Note that although it is possible to introduce additional couplings between MBSs on different sides, the resulting spectrum would differ significantly with respect to the topological case, and therefore, it would be distinguishable. 
For this reason, we employ a specific set of parameters to obtain a cross-correlation spectrum closely resembling that of the two MBSs case, emphasizing the significance of making differentiation possible. Hence, we specifically maintain identical spin angles $\Theta_D=\Theta$ and avoid overlap between MBSs on the same side, i.e. $\xi_D=0$, describing the scenario of coupled zero-energy Andreev bound states. A related model for coupling a single zero-energy Andreev bound state to a lead was studied in Ref. \cite{Haim2015}.

We start the comparison with the cross-correlations spectroscopy in Fig.~\ref{fig:qMBS}(b) and (c), where we observe similar resonance lines with respect to the MBSs case, see Fig.~\ref{fig:MBS}, since the tunneling terms in the qMBSs system also allow for finite triplet resonance lines. However, there are some qualitative differences relative to the cross-correlation resonance lines that allow us to differentiate between MBSs and qMBSs.
Therefore, we investigate line cuts for different tunneling regimes, where we relate the tunneling amplitude $t$ between MBSs and QDs to the tunneling amplitude $t_{AB}$ between MBSs on different sides.

Here, we can also identify broad and sharp resonance peaks, as can be seen from the line cut of the cross-correlations along $\varepsilon_{eA}=\Delta_Z$ for the intermediate tunneling regime ($|t|\approx|t_{AB}|$), see Fig.~\ref{fig:qMBS}(d). In this occasion, the number of both sharp and broad resonances has increased. Moreover, broad peaks show asymmetries and exhibit sharp resonances on top. Again, every resonance corresponds to an anticrossing in the energy spectrum, where the number of anticrossings around $\mu_e=0$ is highly increased. This is because the number of eigenstates is doubled compared to the case of two MBSs, since four MBSs comprise two complex fermions.

In contrast to the previous 4 sharp resonances appearing around $\varepsilon_{eB}=-3\Delta_Z$, here we observe 16 sharp resonances for $S_{\up\up}$, which serves as the upper boundary for four MBSs, since $2^{\# \rm MBSs}=16$. These resonances originate from anticrossings in the spectrum - 8 for the even and 8 for the odd parity. The mechanism for the emission of two correlated photons is similar to the emission cycles given in the MBSs case. But the doubled number of complex fermions gives rise to a fourfold number of anticrossings as well as resonance peaks compared to the case of two MBSs. For instance, we now have four possibilities for the hybridization between even-parity states, involving either both QDs being empty or both being occupied,
\begin{align}
&\ket{0,0,0,0}\leftrightarrow\ket{\up,0,0,\up},\\
&\ket{0,0,0,0}\leftrightarrow\ket{\up,1,1,\up},\\
&\ket{0,1,1,0}\leftrightarrow\ket{\up,0,0,\up},\\
&\ket{0,1,1,0}\leftrightarrow\ket{\up,1,1,\up},
\end{align}
resulting in four resonance peaks.
In contrast, there is only one anticrossing between $\ket{0,0,0}$ and $\ket{\up,0,\up}$ for the case of two MBSs.
The same holds for the anticrossings between states with both QDs occupied or only one being occupied ($\ket{\up,n_c,n_c,\up}\leftrightarrow\ket{\down,n_c',\bar{n}_c',0}$), that lead to other four resonance peaks, thus in total there are 8 peaks for the even parity.

In the case of broad resonances, $S_{\down\down}$ shows asymmetries, and also tends to vanish at resonance at $\varepsilon_{eB}=\Delta_Z$. Unlike the topological case, it exhibits sharp peaks on top. Here, the spectrum exhibits more anticrossings due to the increased number of states, thereby enabling the occurrence of additional resonances. For instance, the peak close to $\varepsilon_{eB}=2\Delta_Z$ results from the emission cycles involving highly hybridized odd-parity states
\begin{align}\label{eq:qMBS_broad}
&\ket{0,1,0,0} \Leftrightarrow \ket{\down,0,0,0} \Leftrightarrow \ket{\down,0,1,\down}\xrightarrow{\text{2 ph}}\ket{0,0,1,0},\\
&\ket{0,1,0,0}\Leftrightarrow\ket{0,1,1,\down}\Leftrightarrow\ket{\down,0,1,\down}\xrightarrow{\text{2 ph}}\ket{0,0,1,0}.
\end{align}
Because of the weak hybridization between $\ket{0,0,1,0}\leftrightarrow\ket{0,1,0,0}$, that is necessary to go back to the highly hybridized states, the corresponding emission peak is small.

Examining different tunneling regimes, we observe for small tunneling amplitudes ($|t|<|t_{AB}|$), see Fig.~\ref{fig:qMBS}(e), a decrease in the number of visible peaks and that these peaks shift closer together. For instance, for $S_{\up\up}$, the peaks are shifted towards $\varepsilon_{eB}=-3\Delta_Z$. Conversely, in the large tunneling regime ($|t|>|t_{AB}|$), see Fig.~\ref{fig:qMBS}(f), the peaks become broader and move away from each other, which makes it difficult to accurately count the number of peaks, as some may be missed. Nevertheless, the number of visible peaks for $S_{\up\up}$ remains larger than four, thereby excluding the possibility of having only two MBSs in the system.

\section{Conclusions}\label{sec:conclusion}

We investigated theoretically signatures of Majorana bound states (MBSs) that appear in the photonic cross-noise spectroscopy in a $pnp$ junction. The system is composed of a topological superconductor (TSC) coherently tunnel coupled on each side to an optically active quantum dot (QD) forming a $pnp$ junction. In this way, when an electron provided by the TSC ($n$) via the MBS tunnels to a QD level with a given spin direction, it can recombine with a hole provided from a normal conducting part ($p$), resulting in the emission of a photon, whose polarization is locked to the electron spin. Thus, the presence of nonlocal superconducting correlations allows for the correlated emission of a photon on each QD. This phenomenon expresses itself as a finite photonic cross-noise, whose polarization reflects the superconducting spin texture. In this way, our detection scheme allows to probe nonlocal spin-triplet superconducting correlations by the direct measurement of the photonic cross-noise polarization. Remarkably, this signature is robust even in the presence of additional uncorrelated quasiparticles, which enhance local emission processes. 

Unfortunately, although nonlocal superconducting triplet correlations are necessary for the presence of MBSs, they are not specific to this system since, as we showed, trivial quasi-MBSs (qMBSs) can also exhibit them due to the presence of spin-orbit coupling, Zeeman field and superconductivity in these systems. For this reason, we analyzed closely the cross-noise spectroscopy and established a correspondence between the number of complex fermions composed of the qMBSs and the number of cross-correlation resonances. In the case of a TSC junction, a pair of MBSs comprises a single complex fermion, and leads to up to $2^{\#\rm MBSs}=4$ sharp resonances arising along a line cut for each spin component. In contrast, in the case of trivial qMBSs, with two complex fermions or 4 MBSs, up to $2^{\#\rm MBSs}=16$ peaks emerge for each spin component of the cross-correlations.

In summary, our detection scheme allows to differentiate between MBSs and qMBSs by counting the number of sharp resonances of one spin component of the cross-correlations. Importantly, this detection method remains effective, even when qMBSs closely mimic the scenario of two MBSs by examine the cross-correlations spectroscopy in the intermediate tunneling regime. Further, the locking of the spin-information of the QD-spin state to the polarization state of the emitted photon that is readily measured by state of the art photodetectors provides a more direct spectroscopic tool of the spin texture of the (nonlocal) superconducting correlations than charge current measurements.

\section*{Acknowledgments}
We acknowledge financial support by the Deutsche Forschungsgemeinschaft (DFG, German Research Foundation) under Germany’s Excellence Strategy – EXC-2123 QuantumFrontiers – 390837967.

\appendix
\section{Derivation of the master equation for photon emission}\label{app:master_equation}
To calculate the dynamics of the system, we use full counting statistics and a Markovian master equation \cite{Flindt2004,Kaiser2007,Emary2007,Sanchez2008,Dominguez2010}. We discussed the refilling of holes in the main text after Eq.~\eqref{eq:hole_coupling}. For the emission of photons, we split the model into an exactly solvable part $H_0=H_e+H_h+H_{\rm ph}$ and the coupling $V_{\rm rec}$ between the system $H_S=H_e+H_h$, which has the eigenvalue equation $H_S\ket{\psi}=E_{\psi}\ket{\psi}$ with eigenstate $\ket{\psi}=\ket{\psi}_e\times\ket{\psi}_h$ and the photon reservoir $H_{\rm ph}$ with eigenvalues $\hbar \omega_k$, see Eqs.~\eqref{eq:H_e,fermion}, \eqref{eq:H_h}, \eqref{eq:H_ph}, and \eqref{eq:V_rec}. The counting fields $\chi_{D\zeta}$ (conjugate variable to the occupation number operator $\hat{n}_{kD\zeta}=a_{kD\zeta}^\dagger a_{kD\zeta}$) count photons emitted from QD $D=A,B$ with polarization $\zeta=L,R$. 

For the dynamics of the full system, we use the von Neumann equation
\begin{align}\label{eq:vonNeumann}
\partial_t \rho (t) = -\frac{i}{\hbar} [(H_0+V_{\rm rec}),\rho(t)],
\end{align}
where $\rho(t)$ is the density matrix of the full system $H_0+V_{\rm rec}$.
We add the counting fields with a Fourier transformation
\begin{align}\label{eq:FT}
\rho(\chi_{D\zeta},t)=\rho(t)\exp(i \sum_{\hat{n}_{kD\zeta}} \chi_{D\zeta} \hat{n}_{kD\zeta}).
\end{align}
By inserting Eq.~\eqref{eq:FT} into Eq.~\eqref{eq:vonNeumann} we can derive a generalized von Neumann equation,
\begin{align}
\begin{split}
\partial_t\rho(\chi_{D\zeta},t)=&-\frac{i}{\hbar}(H^+(\chi_{D\zeta})\rho(\chi_{D\zeta},t)\\&-\rho(\chi_{D\zeta},t)H^-(\chi_{D\zeta})),
\end{split}
\end{align}
where the Hamiltonian is given by
\begin{align}
\begin{split}
H^\pm (\chi_{D\zeta})=H_0+(g \sum_{k} d_{D\zeta} h_{D\bar{\zeta}}a^\dagger_{kD\zeta} e^{\pm i \chi_{D\zeta}/2}+ \rm H.c.).
\end{split}
\end{align}
Thus, the time evolution of the density matrix is given by 
\begin{align}
\rho(\chi_{D\zeta},t)=e^{-i H^+ (\chi_{D\zeta})t/\hbar}\rho(\chi_{D\zeta})e^{i H^- (\chi_{D\zeta})t/\hbar}.
\end{align}
Now, we transform to the interaction picture, where operators $\hat{O}$ and the density matrix are described by
\begin{align}
&\hat{O}_I(\chi_{D\zeta},t)=e^{i H_0 t/\hbar} \hat{O}(\chi_{D\zeta}) e^{-i H_0 t/\hbar},\\
&\rho_I(\chi_{D\zeta},t)=e^{i H_0 t/\hbar} \rho(\chi_{D\zeta},t) e^{-i H_0 t/\hbar},
\end{align}
respectively, and the time-derivative of the density matrix simplifies to 
\begin{align}\label{eq:rho_interaction}
\begin{split}
\partial_t \rho_I(\chi_{D\zeta},t)=&-\frac{i}{\hbar} V^+_I(\chi_{D\zeta},t)\rho_I(\chi_{D\zeta},t)\\&+\frac{i}{\hbar}\rho_I(\chi_{D\zeta},t) V^-_I(\chi_{D\zeta},t).
\end{split}
\end{align}
Here, only the interaction
\begin{align}
V^\pm_I(\chi_{D\zeta},t)&= \sum_k g e^{\pm i\chi_{D\zeta}/2} d_{D\zeta}(t)h_{D\bar{\zeta}}(t)a_{kD\zeta}^\dagger(t)+\rm H.c.\\
&= S_{D\zeta}(t)P^\dagger_{\pm}(\chi_{D\zeta},t)+\rm{H.c.}
\end{align}
appears, where we introduce two new operators $S_{D\zeta}(t)=d_{D\zeta}(t)h_{D\bar{\zeta}}(t)$ and $P_\pm(\chi_{D\zeta},t)=\sum_k g^* e^{\mp i\chi_{D\zeta}/2}a_{kD\zeta}(t)$, which obey the commutation relation $[S_{D\zeta}(t),P_{\pm}(\chi_{D\zeta},t)]=0$.
Integrating Eq.~\eqref{eq:rho_interaction} and inserting it again, leads to
\begin{align}\label{eq:int_rho}
\begin{split}
\partial_t \rho_I(\chi_{D\zeta},t)=&\\
-\frac{i}{\hbar} V^+_I(\chi_{D\zeta}&,t)\rho_I(\chi_{D\zeta},t)
+\frac{i}{\hbar}\rho_I(\chi_{D\zeta},t) V^-_I(\chi_{D\zeta},t)\\
+\frac{1}{\hbar^2}\int_{0}^t dt' [
&-V_I^+(\chi_{D\zeta},t) V_I^+ (\chi_{D\zeta},t')\rho_I (\chi_{D\zeta},t')\\
&+V_I^+(\chi_{D\zeta},t)\rho_I(\chi_{D\zeta},t')V_I^-(\chi_{D\zeta},t')\\
&+V_I^+(\chi_{D\zeta},t')\rho_I(\chi_{D\zeta},t')V_I^-(\chi_{D\zeta},t)\\
&-\rho_I(\chi_{D\zeta},t')V_I^-(\chi_{D\zeta},t')V_I^-(\chi_{D\zeta},t)].
\end{split}
\end{align}
In the further calculation, we neglect the reaction from the photon reservoir to the system, such that the total density matrix can be written as
\begin{align}
\rho_I(\chi_{D\zeta},t)=\rho_{SI}(\chi_{D\zeta},t)\otimes\rho_{\rm{ph}},
\end{align}
and trace out the photon reservoir, which leads to $\rho_{SI}(\chi_{D\zeta},t)={\rm Tr_{ph}}[\rho_I(\chi_{D\zeta},t)]$,
\begin{align}
\braket{P^{(\dagger)}_\alpha(\chi_{D\zeta},t)}=0,
\end{align}
and the correlators
\begin{align}
&\braket{P_\alpha(\chi_{D\zeta},\tau)P_\beta(\chi_{D\zeta})}=\braket{P^\dagger_\alpha(\chi_{D\zeta},\tau)P^\dagger_\beta(\chi_{D\zeta})}=0,\label{eq:correlator=0}\\
&\braket{P^\dagger_\alpha(\chi_{D\zeta},\tau)P_\beta(\chi_{D\zeta})}=|g|^2 e^{(\alpha-\beta)i\chi_{D\zeta}/2}\sum_k e^{i\tau\omega}\braket{\hat{n}_{kD\zeta}},\\
\begin{split}&\braket{P_\alpha(\chi_{D\zeta},\tau)P^\dagger_\beta(\chi_{D\zeta})}\\  &\quad =|g|^2 e^{-(\alpha-\beta)i\chi_{D\zeta}/2}\sum_k e^{-i\tau\omega}(1+\braket{\hat{n}_{kD\zeta}}),
\end{split}
\end{align}
with $\alpha,\beta=+,-$, $\tau=t-t'$, and $\braket{\hat{n}_{kD\zeta}}=\braket{a_{kD\zeta}^\dagger a_{kD\zeta}}={\rm Tr}_{\rm ph}[a_{kD\zeta}^\dagger a_{kD\zeta}\rho_{\rm ph}]$.
We assume that no optical photons are present in equilibrium, so $\braket{\hat{n}_{kD\zeta}}=0$.

Now, we use the Markov approximation, such that $\rho_{SI}(\chi_{D\zeta},t')\to\rho_{SI}(\chi_{D\zeta},t)$ and extend the integral in Eq.~\eqref{eq:int_rho} to infinity. We take all matrix elements in the system state, perform the integration over $\tau$ and go back to the Schrödinger picture, where we obtain in secular approximation
\begin{align}
\begin{split}
&\partial_t\rho_{S}^{\psi\psi}(\chi_{D\zeta},t)=\frac{2\pi}{\hbar}\sum_k \sum_{\psi'}|g|^2\\&(-|\bra{\psi'}S_{D\zeta}\ket{\psi}|^2 \rho_{S}^{\psi\psi}(\chi_{D\zeta},t)\delta(E_\psi-E_{\psi'}-\hbar\omega_k)\\&+e^{i\chi_{D\zeta}}|\bra{\psi}S_{D\zeta}\ket{\psi'}|^2\rho_{S}^{\psi'\psi'}(\chi_{D\zeta},t)\delta(E_{\psi'}-E_\psi-\hbar\omega_k)).
\end{split}
\end{align}
We integrate over k ($\sum_k\to\frac{2\pi}{L}\int dk$) to obtain Eq.~\eqref{eq:mastereq}.

\section{Spinless model}\label{app:spinless}

\begin{figure*}
\centering
	\includegraphics[width=\textwidth]{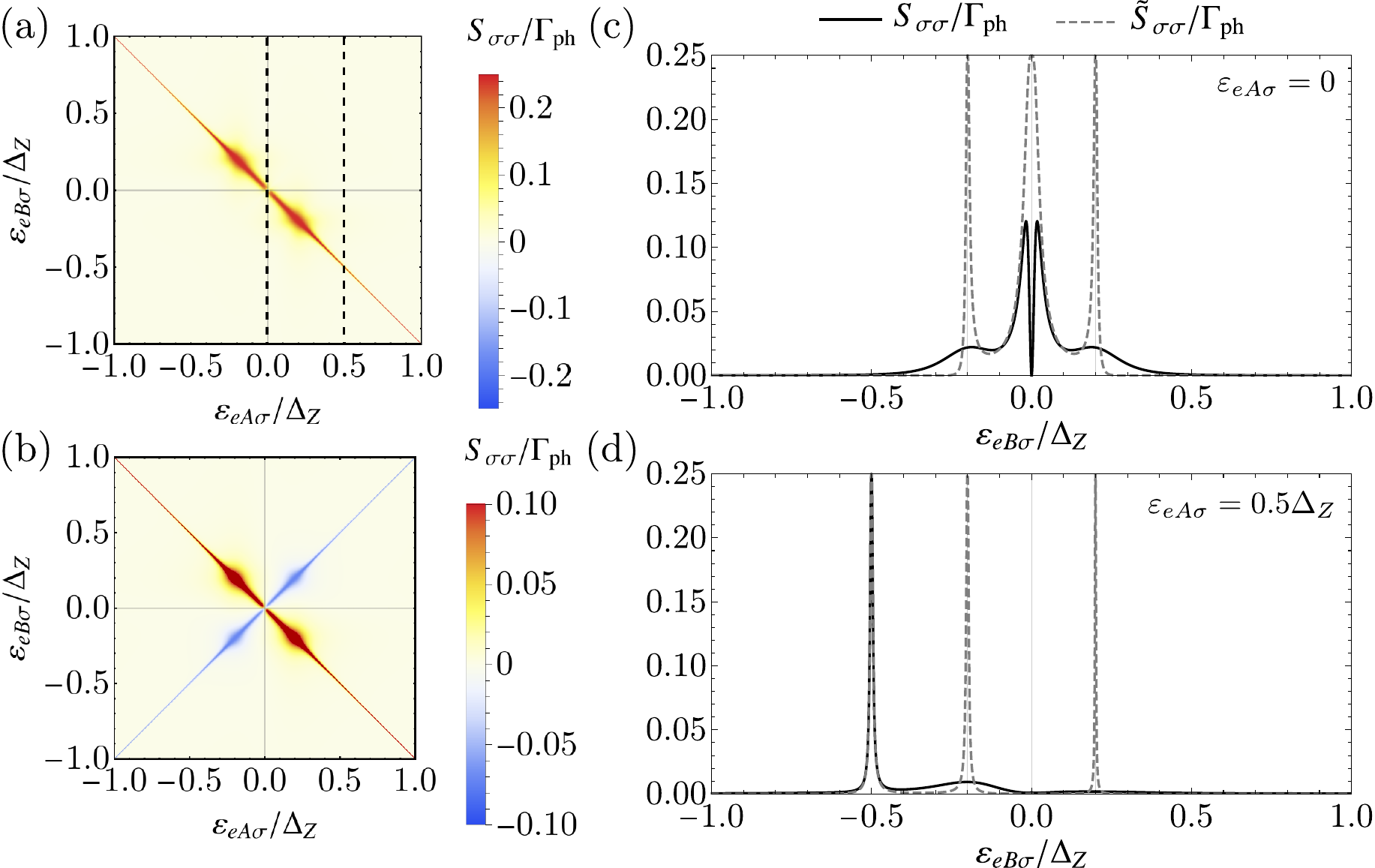}
	\caption{Cross-correlations spectroscopy for the spinless model. We show $S_{\sigma\sigma}$ as a function of the QD energies $\varepsilon_{eA\sigma}$ and $\varepsilon_{eB\sigma}$ in the small tunneling regime ($|t_{D\sigma}|<|\xi|$) for $\Gamma_{\rm qp}=0$ (a) and $\Gamma_{\rm qp}=\Gamma_{\rm ph}/2$ (b). Additionally, we show line cuts of $ S_{\sigma\sigma}$ from panel (a) (solid line) and the analytical expression $\tilde S_{\sigma\sigma}$ (dashed line) along $\varepsilon_{eA\sigma}=0$ (c) and $\varepsilon_{eA\sigma}=0.5\Delta_Z$ (d). The parameters are $\xi=0.2\Delta_Z$ and $t_{A\sigma}=it_{B\sigma}=t=0.03\Delta_Z$.}
	\label{fig:spinless}
\end{figure*}

To get a further inside of the underlying processes in the cross-correlations spectroscopy, we reduce the system to a spinless model, that is applicable, when the Majorana spin points in $\pm z$ direction ($\Theta=0,\pi$) or when $|t_{D\sigma}|\ll|\Delta_Z|$. By replacing Eq.~\eqref{eq:H_e,fermion} with
\begin{align}\label{eq:H_spinless}
\begin{split}
H_{e\sigma}&=\sum_{D}\varepsilon_{eD\sigma} d_{D}^\dagger d_{D}+\xi\left(
\hat{n}_c-\frac{1}{2}\right)\\
&+\left[t_{A\sigma} d_{A} (c^\dagger +c) + it_{B\sigma} d_{B}( c^\dagger-c) +\text{H.c.}\right],
\end{split}
\end{align}
where $\sigma$ only appears as an index, we can investigate the cross-correlation $S_{\sigma\sigma}$, see Fig.~\ref{fig:spinless}.
The spinless model resembles nonlocal triplet correlations for parallel spins, as it leads to a positive resonance line along $\varepsilon_{eA\sigma}=-\varepsilon_{eB\sigma}$, see Fig.~\ref{fig:spinless}(a), where the CAR condition is fulfilled. Additionally, when uncorrelated quasiparticles tunnel onto the QDs ($\Gamma_{\rm qp}>0$), the cross-correlation remains robust and a negative resonance line emerges along $\varepsilon_{eA\sigma}=\varepsilon_{eB\sigma}$ due to ECT processes, see Fig.~\ref{fig:spinless}(b).

The emergence of positive triplet correlations stems from the emission of correlated photon pairs. For parallel spins, this process can effectively be described by a triplet pairing on two QDs,
\begin{align}\label{eq:SC}
H_{SC}= \sum_D \varepsilon_{eD\sigma} d_D^\dagger d_D + (\Delta_p d_A^\dagger d_B^\dagger + \rm H.c.),
\end{align}
where $\Delta_p\in\mathbb{R}$ is the pairing amplitude of the effective superconductivity on the QDs due to the proximity effect. Note that we effectively traced out the TSC hosting the MBSs. We can diagonalize Eq.~\eqref{eq:SC} in the basis $\ket{n_A,n_B}$, where $\hat{n}_D=d_D^\dagger d_D$ is the occupation number operator, and obtain an analytical expression for the effective cross-correlations
\begin{align}
&\tilde S_{\sigma\sigma}=\Gamma_{\rm ph}\frac{2 \Delta_p^2\big[2 \Delta_p^2 +(\varepsilon_{eA\sigma}+\varepsilon_{eB\sigma})^2\big]}{\big[4 \Delta_p^2+(\varepsilon_{eA\sigma}+\varepsilon_{eB\sigma})^2\big]^2},\label{eq:noisecross_SC}
\end{align}
by using Eq.~\eqref{eq:crossnoise} accordingly.

To include the contribution of the MBSs, we perform a Schrieffer-Wolff transformation \cite{Schrieffer1966} on Eq.~\eqref{eq:H_spinless} in the limit $|t_{D\sigma}|/|\varepsilon_{eD\sigma}-\xi|\ll 1$, so the occupation of the nonlocal fermion only enters virtually. We can split Eq.~\eqref{eq:H_spinless}, in an unperturbed part $H^0=\sum_{D=A,B}\varepsilon_{eD\sigma} d_{D}^\dagger d_{D}+\xi\left(
\hat{n}_c-\frac{1}{2}\right)$ and a small perturbation $H'=\left[t_{A\sigma} d_{A} (c^\dagger +c) + it_{B\sigma} d_{B}( c^\dagger-c) +\text{H.c.}\right]$. We separate the eigenstates of $H^0$ in low energy states $\ket{m}=\ket{\psi}_e$ for $\hat{n}_c=0$ with eigenenergy $E_m$ and high energy states $\ket{l}=\ket{\psi}_e$ for $\hat{n}_c=1$ with eigenenergy $E_l$, i.e. we assume the occupied nonlocal fermion as the high energy sector.
The matrix elements of the effective Hamiltonian up to second order are given by \cite{Winkler2003}
\begin{widetext}
\begin{align}
\begin{split}
	\tilde{H}_{mm'}=H^0_{mm'}+H'_{mm'}+\frac{1}{2}\sum_l H'_{ml}H'_{lm'} \times[\frac{1}{E_m-E_l}+\frac{1}{E_{m'}-E_l}]+\mathcal{O}(H'^3).
	\end{split}
\end{align}
We obtain the effective Hamiltonian for the even and odd parity,
\begin{align}
\begin{split}\label{eq:SWT_even}
\tilde{H}_{\rm even}&=\left[\sum_D \left(\varepsilon_{eD\sigma}+ \frac{|t_{D\sigma}|^2}{\varepsilon_{eD\sigma}-\xi}\right)-\frac{\xi}{2}\right]\hat{n}_A \hat{n}_B
-\left[\sum_D \left(\frac{|t_{D\sigma}|^2}{\varepsilon_{eD\sigma}+\xi}\right)+\frac{\xi}{2}\right] (1- \hat{n}_A \hat{n}_B)\\
&+\left[i t_{A\sigma} t_{B\sigma} \left(\frac{\xi}{\varepsilon_{eA\sigma}^2-\xi^2}+\frac{\xi}{\varepsilon_{eB\sigma}^2-\xi^2}\right)d_B d_A+\rm H.c.\right],
\end{split}
\end{align}
\begin{align}
\begin{split}
\tilde{H}_{\rm odd}&=\sum_D \left(\varepsilon_{eD\sigma}+ \frac{|t_{D\sigma}|^2}{\varepsilon_{eD\sigma}-\xi}-\frac{|t_{\bar{D}\sigma}|^2}{\varepsilon_{e\bar{D}\sigma}+\xi}-\frac{\xi}{2}\right)d_{D}^\dagger d_{D}
+\left[i t_{A\sigma}^* t_{B\sigma} \left(\frac{\xi}{\varepsilon_{eA\sigma}^2-\xi^2}+\frac{\xi}{\varepsilon_{eB\sigma}^2-\xi^2}\right)d_A^\dagger d_B+\rm H.c.\right],
\end{split}
\end{align}
\end{widetext}
respectively. The effective pairing amplitude $\tilde\Delta_{p}$ in
Eq.~\eqref{eq:eff_pairing} can be extracted from Eq.~\eqref{eq:SWT_even}, which accounts for the pairing
between $\ket{0,0}$ and $\ket{\sigma,\sigma}$.
Thus, we can deduce an effective pairing amplitude
\begin{align}\label{eq:eff_pairing} 
\tilde{\Delta}_p = i t_{A\sigma}t_{B\sigma} \left(\frac{\xi}{\varepsilon_{eA\sigma}^2-\xi^2}+\frac{\xi}{\varepsilon_{eB\sigma}^2-\xi^2}\right).
\end{align}
which we can insert into Eq.~\eqref{eq:noisecross_SC} to obtain the analytical expression for the cross-correlations $\tilde{S}_{\sigma\sigma}$ in Eq.~\eqref{eq:S_ana}.

In Figs.~\ref{fig:spinless}(c) and (d), we investigate line cuts of $S_{\sigma\sigma}$ along $\varepsilon_{eA\sigma}=0$ and $\varepsilon_{eA\sigma}=0.5\Delta_Z$, respectively, and compare them to the effective $\tilde S_{\sigma\sigma}$. When one QD level is in resonance with $\mu_e=0$, the cross-correlations $S_{\sigma\sigma}$ exhibit four peaks - two at $\varepsilon_{eB\sigma}=\pm\xi$ and two close to resonance at $\varepsilon_{eB\sigma}=0$. In this scenario, the analytical prediction matches the width of the peaks at resonance near $\varepsilon_{eB\sigma}=0$. However, a discrepancy arises in the peak heights, which is given by $\Gamma_{\rm ph}/4$ in the analytical model. Additionally, that $S_{\sigma\sigma}$ vanishes at $\varepsilon_{eB\sigma}=0$ is due to the fact that CAR and ECT processes compensate each other, a feature not captured by the analytics since ECT processes are not included. Traces of divergencies in the analytical model appear at $\varepsilon_{eB\sigma}\pm\xi$, aligning with the positions of the small broad peaks at $\varepsilon_{eB\sigma}\pm\xi$ from the effective model. Conversely, when the QD levels are detuned from resonance at $\mu_e=0$, the cross-correlations exhibit a sharp peak, satisfying the CAR condition ($\varepsilon_{eA\sigma}=-\varepsilon_{eB\sigma}$) as predicted by the analytical model. Additionally, observable flat peaks at $\varepsilon_{eB\sigma}\pm\xi$ coincide with traces of divergencies from the analytical model.

\begin{figure}[t!]
	\includegraphics[width=\columnwidth]{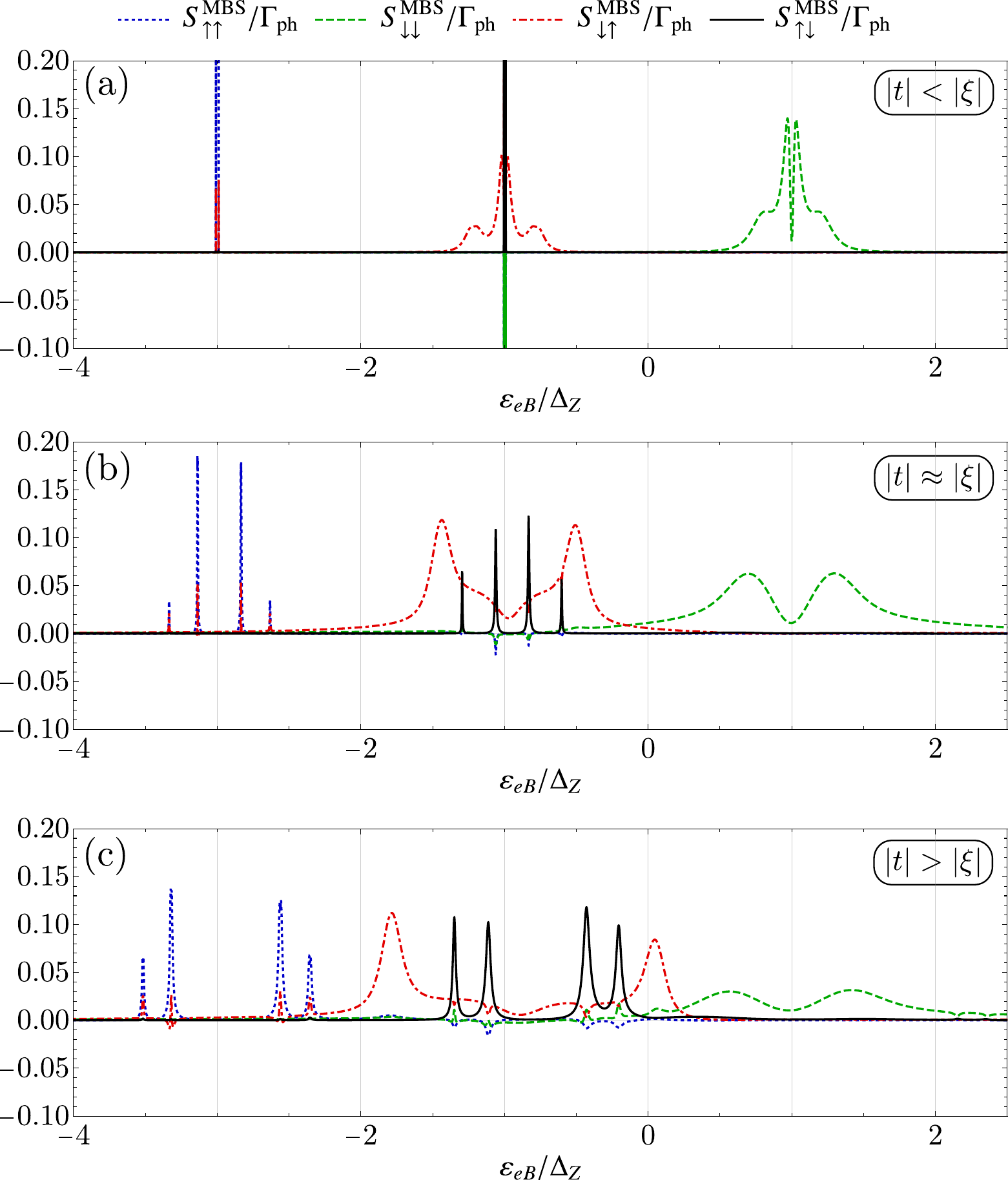}
	\caption{Cross-correlations spectroscopy for MBSs including nonlocal tunneling amplitudes. We show $S_{\zeta\zeta'}$ along $\varepsilon_{eA}=\Delta_Z$ as a function of $\varepsilon_{eB}$ for (a) small, (b) intermediate and (c) large tunneling regimes. The parameters are $t_{nl}=0.16 t$, $\xi=0.2\Delta_Z$, $\Theta=0.75\pi$, $\Theta_{nl}=\pi/2$, and $\Gamma_{\rm qp}=0.$ For the tunneling amplitude we use $t=0.05\Delta_Z$ in (a), $t=0.25\Delta_Z$ in (b), and $t=0.5\Delta_Z$ in (c).}
	\label{fig:MBSnl}
\end{figure}
\section{Influence of nonlocal tunnel couplings between MBSs and QDs}\label{app:MBSnl}

In this section, we investigate the influence of nonlocal tunneling amplitudes, i.e. the tunneling between QD levels and MBSs on opposite sides \cite{Prada2017, Schuray2017,Clarke2017}. To include the nonlocal tunneling terms, we replace the second line in Eq. \eqref{eq:H_e,fermion} by
\begin{align}
\begin{split}
H_{nl}&=\sum_{D,\sigma}[(t_{D1\sigma}+i t_{D2\sigma}) d_{D\sigma}c^\dagger\\
&+ (t_{D1\sigma}-i t_{D2\sigma}) d_{D\sigma} c +\text{H.c.}],
\end{split}
\end{align}
where $t_{Di\sigma}$ is the tunneling amplitude between electrons with spin $\sigma$ on QD $D$ and MBS $\gamma_i$, where we set the new notation $\gamma_A=\gamma_1$ and $\gamma_B=\gamma_2$, compared to Eq.~\eqref{eq:H_e}. 
We can parametrize the local and nonlocal tunneling amplitudes by the angles $\Theta$ and $\Theta_{nl}$, respectively,
\begin{align}
\begin{split}
&t_{A1\uparrow}=-i t_{B2\uparrow}=t\cos(\Theta/2),\\
&t_{A1\downarrow}=i t_{B2\downarrow}=t\sin(\Theta/2),\\
&t_{B1\uparrow}=-i t_{A2\uparrow}=t_{nl}\cos(\Theta_{nl}/2),\\
&t_{B1\downarrow}=i t_{A2\downarrow}=t_{nl}\sin(\Theta_{nl}/2),
\end{split}
\end{align}
with tunneling amplitudes $t$ and $t_{nl}$. We can estimate the ratio of the tunneling amplitudes for overlapping MBSs with $t_{nl}/t=0.16$ adapted from Ref. \cite{Deng2018}.

In Fig.~\ref{fig:MBSnl}, we show the results of the cross-correlations spectroscopy for three different tunneling regimes. Compared to the model without nonlocal couplings, see Fig.~\ref{fig:MBS}, we observe no qualitative differences. Note that a higher ratio could potentially lead to the emergence of additional resonance peaks, which could be erroneously associated with trivial states.


%

\end{document}